
\magnification= \magstep 1
\global\newcount\meqno 
\def\eqn#1#2{\xdef#1{(\secsym\the\meqno)} 
\global\advance\meqno by1$$#2\eqno#1$$} 

\def\WR{D. Weaire and N. Rivier, {\it Contemp. Phys.}  {\bf 25} (1984) 59.} 
\def\STA{J. Stavans, {\it Rep. Prog. Phys.} {\bf 54} (1993) 733.}
\def\SCHU{H. G. Schuster, {\it Deterministic Chaos}  (Physik-Verlag, Weinheim 
1984).}
\def\LAV{P. M. Oliveira, M. A. Continentino and E. V. Anda, {\it Phys. Rev. B} {\bf 29} 
(1984) 2808. 
B. K. Southern, A. A. Kumar, P. D. Loly and M-A. S. Tremblay, {\it Phys. Rev. B} {\bf 27} 
(1983) 1405.
D. A. Lavis, B. W. Southern and S. G. Davinson, {\it J. Phys. C} {\bf 18} (1985) 
1387.}
\def\KREY{see e.g. E. Kreyszig, {\it Differential Geometry} (Dover, New York, 
1991).}
\def\K{W. Thomson (Lord Kelvin, {\it Phil. Mag. } {\bf 24} (5) (1887) 503.}
\def\WIL{R. Williams, {\it The Geometrical Foundation of Natural Structure} 
Dover, New York, 1979).}  
\def\COX{H.M.S. Coxeter, {\it Regular Polytopes} (Dover, New York, 1973).} 
\def\GLAZ{J. A. Glazier and D. Weaire, {\it Phil. Mag. Lett.} {\bf 70} (1994) 351.} 
\def\WILS{R. E. Williams, {\it Science} {\bf 161} (1968) 276.}
\def\POL{B. R. Pollard, {\it An Introduction to Algebraic Topology} (University of 
Bristol Press 1977).}
\def\CHE{S. S. Chern, {\it Ann. Math.} {\bf 45} (1944) 747.}  
\def\WEG{M. Goldberg, {\it Tohoku Math. J.} {\bf 40} (1934) 226.}
\def\FRK{F.C. Frank and J.S. Kasper, {\it Acta crystallogr.} {\bf 12} (1959) 483.}
\def\SHO{D.P. Shoemaker and C.B. Shoemaker, {\it Acta crystallogr.} {\bf B42} 
(1986) 3.}
\def\WP{D. Weaire and R. Phelan, {\it Phil. Mag. Lett.} {\bf 69} (1994) 107.} 
\def\RIV{N. Rivier, {\it Phil. Mag. Lett.} {\bf 69} (1994) 297.} 
\def\WEDIS{D. Weaire, private communication.} 
\def\SM{J.F. Sadoc and R. Mosseri, {\it J. Phys. (Paris)} {\bf 46} (1985) 1809.}
\def\KU{R. Kusner, {\it Proc. R. Soc. Lond. A } {\bf 439} (1992) 683.}
\def\AsRi{T. Aste and N. Rivier, {\it J. Phys. A} {\bf 28} (1995) 1381.}
\def\RILI{N. Rivier and A. Lissowski, {\it J. Phys. A } {\bf 15} (1982) L143.}
\def\W{D. Weaire and R. Phelan, {\it Phil. Mag. Lett.} {\bf 70} (1994) 351.}
%
\global\newcount\refno
\def\ref#1{\xdef#1{[\the\refno]}
\global\advance\refno by1#1}
\global\refno = 1
\hoffset =0.8 true cm
\hsize =	15.8 true cm
\vsize = 21 true cm
\tolerance 10000

%

\baselineskip=0.1cm
\baselineskip 12pt plus 1pt minus 1pt
\vskip .5in
\centerline{\bf FROM ONE CELL TO THE WHOLE FROTH: A DYNAMICAL 
MAP}
\vskip 24pt
\centerline{T. Aste \footnote{\S}{On leave from C.I.I.M., Universit\'a di Genova, 
Genova, Italy}, D. Boos\'e and N. Rivier}
\vskip 12pt
\centerline{\it Laboratoire de Physique Th\'eorique, Universit\'e Louis Pasteur}
\centerline{\it 67084\ \ Strasbourg, France}
\vskip 1.cm
\vfill
\baselineskip 12pt plus 1pt minus 1pt
\centerline{{\bf ABSTRACT}}

We investigate two and three-dimensional shell-structured-inflatable froths, which 
can be constructed by a recursion procedure adding successive layers of cells around 
a germ cell. 
We prove that any froth can be reduced into a system of concentric shells.
There is only a restricted set of local configurations for which the recursive inflation 
transformation is not applicable. 
These configurations are inclusions between successive layers and can be treated as 
vertices and edges decorations of a shell-structure-inflatable skeleton.
The recursion procedure is described by a logistic map, which provides a natural 
classification into Euclidean, hyperbolic and elliptic froths. 
Froths tiling manifolds with different curvature can be classified simply by 
distinguishing between those with a bounded or unbounded number of elements per 
shell, without any a-priori knowledge on their curvature.
A new result, associated with maximal orientational entropy, is obtained on 
topological properties of natural cellular systems. 
The topological characteristics of all experimentally known tetrahedrally close-
packed structures are retrieved.

\medskip
\vfill
\centerline{\bf I. INTRODUCTION}
\medskip
\nobreak
\xdef\secsym{1.}\global\meqno = 1
\medskip
A froth is a (topologically stable) division of space by cells, which are convex 
polytopes (polygons in 2D, polyhedra in 3D) of various shapes and sizes. These 
geometrical systems have  attracted much attention in recent years, both theoretically 
and experimentally \ref\WRr, \ref\STAr. 
The aim in this work is to study a specific class of froths, namely those which are 
reducible to a set of concentric shells. 
These particular froths are structured as if constructed in the following way. In a first 
stage, cells 
are added to a germ cell, forming around it a first layer whose external surface 
constitutes the second shell.
In a second stage, cells are added to the first shell so as to form a second layer of 
cells encircling the first one, and so on. 
We emphasize that the words ``germ'' and ``stage'' are purely pictorial and do not 
imply any particular mode of growth since any cell of a generic shell-structured froth 
may play the role of its germ cell. Such a froth is called  shell-structured-inflatable 
from now on. 

A definition of a shell-structured-inflatable froth requires the notion of topological 
distance between cells. The topological distance $t$ between two cells $A$ and $B$ 
is defined as the smallest number of edges crossed by a path connecting $A$ and 
$B$. 
The germ cell is therefore at distance $t=0$.
A shell $(t)$ is defined as the interface between two sets of cells distant by $t$ and 
$t+1$ from the germ cell. A 2D froth is {\it shell-structured-inflatable} if it satisfies 
the following two conditions :

\noindent
1) For any set cells, equidistant to the germ cell, there exists a closed non self-
intersecting path which goes only through these cells and connects all of them.

\noindent 
2)  Any cell at distance $t$ from the germ cell is the neighbour of at least one cell at 
distance $t+1$.

\noindent 
Two consecutive shells $(t)$ and $(t+1)$ of a shell-structured-inflatable froth are 
connected through a set of disjoint edges with one vertex on shell $(t)$ and the other 
on shell $(t+1)$.
These two shells are closed loops of edges delimiting the layer $(t+1)$ of cells which 
are at distance $t+1$ from the germ cell. 
Shell $(t)$ divides the froth into an internal froth, constituted of cells at distances $r 
\le t$, and an external froth, with cells at distances $r>t$.
The extension to a 3D shell-structured-inflatable froths is straightforward and is 
given in Appendix B.2. 

In  this paper we prove that the 2D and 3D shell-structured-inflatable froths  are 
constructed according to a recursion procedure which is the logistic map \ref\SCHUr, 
well-known in the theory of dynamical systems. 
The logistic map provides a natural classification of these froths according to the 
behaviour of the number of edges per shell as the topological distance $t$ increases.

Any given froth is  not necessarily shell-structured-inflatable. 
However, it has to be noted that a froth can always be decomposed into shells with 
respect to an arbitrarily chosen germ cell. 
In this decomposition, each cell of the layer $(t)$ belongs to one of two categories. 
The cells of the first category, individually, have neighbours in both layers $(t-1)$ 
and $(t+1)$ and, collectively, are building up a complete ring around the chosen 
germ cell. The set of all these rings constitutes the ``skeleton'' of the shell-structure. 
The cells of the second category have neighbours in only one of the two layers $(t-
1)$ or $(t+1)$.
These cells can be considered as local topological defects included between the rings 
of the ``skeleton'' of the shell-structure. 
The ``skeleton'' is itself a space-filling froth which is shell-structured-inflatable. 
The recursion procedure that we are studying applies to such a structure.

The plan of this paper is the following. 
In Section 2, we derive the recursion procedure associated with 2D shell-structured-
inflatable froths and show that it can be written as the logistic map. 
The resulting classification into Euclidean, hyperbolic and elliptic froths is 
discussed. 
In Section 3, it is  shown that the recursion procedure in 3D is again described by a 
logistic map. The curvature of the embedding space is classified as for the 2D froths. 
Section 4 gives examples of space-filling cellular structures which fit into the 
classification of 3D shell-structured-inflatable froths provided by the logistic map. 
In Section 5, a new bound on topological properties of natural cellular structures is 
obtained.  The topological properties of all experimentally known tetrahedrally 
close-packed (t.c.p.) structures are retrieved under the hypothesis of shell-
reducibility. 
A conclusion emphasizes the main results of the paper. 
In Appendix A, the recursion procedure is generalized to 2D shell-structured-
inflatable networks with vertex coordination larger than 3. 
Local topological defects in 2 and 3D shell-reducible but not inflatable froths are 
considered in Appendix B. 
Random 3D Euclidean froths are constructed from 2D random shell networks in 
Appendix C.

\bigskip
\centerline{\bf II. RECURSION PROCEDURE FOR 2D FROTHS} 
\nobreak
\medskip
\nobreak
\xdef\secsym{2.}\global\meqno = 1
\medskip
\nobreak
This section is concerned with two dimensional shell-structured-inflatable froths. 
The recursion procedure is derived here for froths and  it is extended to networks 
with vertex coordination larger than 3 in Appendix A. Fig.(1) shows an example of a 
froth with the various shells indicated by bold lines and labelled by the index $t$ 
(the shell $t = 0$ corresponding to the boundary of the germ cell). 
Let $V_{+(-)}^{(t)}$ be the number of vertices going out from shell $(t)$ towards 
shell $(t+1)$ (towards shell $(t-1)$). 
Let $F^{(t)}$ be the number of cells in the layer between shells $(t)$ and $(t+1)$. 
If $\langle n\rangle$ is the average number of edges per cell in layer $(t)$, the edges 
in this layer are accounted for, as follows
\eqn\iterI{ \langle n\rangle F^{(t)}=V_-^{(t)}+2V_+^{(t)}+ 2V_-
^{(t+1)}+V_+^{(t+1)} \;\;\;\; .}
In the right-hand-side of this equation, the quantity $V_-^{(t)}+V_+^{(t)}$  is the 
total number of vertices constituting shell $(t)$, the quantity $V_-
^{(t+1)}+V_+^{(t+1)}$ is the total number of vertices constituting 
shell $(t+1)$ whereas the quantity $V_-^{(t+1)}+V_+^{(t)}$ gives the number of 
vertices (counted twice) bounding the edges separating the cells comprised between 
shells $(t)$ 
and $(t+1)$. Since $V_+^{(t)}=V_-^{(t+1)}$ and $F^{(t)}=V_+^{(t)}$, one has the 
recursion equation
\eqn\iterII{ \langle n\rangle V_+^{(t)}=4V_+^{(t)}+V_+^{(t+1)}+V_-^{(t)} \;\;\;\; . 
}
The matrix form of this recursion equation is
\eqn\TI{ \Bigg(\eqalign{ &V_+^{(t+1)} \cr &V_-^{(t+1)} } \Bigg)= \Bigg(\eqalign{ 
&s \;\;\;\;-1 \cr &1 \;\;\;\;\;\;\;0} \Bigg)
\Bigg(\eqalign{ &V_+^{(t)} \cr &V_-^{(t)} } \Bigg) \;\;\; ,}
with recursion parameter $s=\langle n \rangle - 4$. 
Eq.\TI\ generates recursively the whole froth from the germ cell.
In general, the quantity $\langle n \rangle$ changes from one layer to the next. 
Hence the recursion parameter should depend on the distance $t$. 
However, the value of $\langle n \rangle$ associated to a layer of cells at a distance 
$t$ from the germ cell must, as $t \rightarrow \infty$, converge to the average value 
for any cell in the froth. 
Moreover, since the choice of the germ cell is completely arbitrary, the quantity 
$\langle n \rangle$ associated with layer $(t)$ is an average. 
Consequently, the recursion parameter can be taken as an effective quantity which is 
independent of $t$, and the quantity $\langle n \rangle$ is then the average number 
of edges per cell in the froth. 
The initial conditions in Eq.\TI\ are then $V_-^{(0)}=0$ and $V_+^{(0)}=\langle n 
\rangle$. 

The recursion procedure described by Eq.\TI\ appears also in other instances, such as 
in the computation by decimation of the electronic energy spectrum in the 1D tight-
binding model \ref\LAVr. In this case, the variables $V^{(t)}$ are replaced by the 
components of the electronic wave-functions in the basis of the site states, and  the 
recursion parameter $s$ is the (dimensionless) energy of the electron.

Eq.\TI\ gives an immediate link between the shell-structured-inflatable froths and the 
logistic map. Indeed, from the relations $sV_+^{(t)}=V_+^{(t+1)}+V_+^{(t-1)}$, 
$sV_+^{(t+1)}=V_+^{(t+2)}+V_+^{(t)}$, and $sV_+^{(t-1)}=V_+^{(t)}+V_+^{(t-
2)}$, one gets 
\eqn\rex{ s_1 V_+^{(t)}=V_+^{(t+2)}+V_+^{(t-2)} \;\;\;\; ,}
with
\eqn\recce{ s_1=s^{2}-2 \;\;\; ,}
and a similar relation for the $V_-$'s. Iterating $j$ times, one obtains  
\eqn\fine{ s_j V_+^{(t)}=V_+^{(t+2^j)}+V_+^{(t-2^j)} \;\;\;\; ,}
with
\eqn\racx{s_{j+1}=s_{j}^{2}-2 \;\;\;\; ,}
and $s_0=s$.
Eq.\racx\ is the trace map of the recursion matrix in Eq.\TI. It is 
a logistic map \SCHUr , with  two (unstable) fixed points $s^{*}=2$ and $s^{*}=-
1$. 
The logistic map decomposes the axis of values of the recursion parameter $s$ into 
two different regions. 
Any point in the region $|s| > 2$ is sent towards infinity by the successive iterations 
of the logistic map. By contrast, if $|s| < 2$, successive iterations of the logistic map 
remain all within this interval. The existence of these two intervals classifies all 2D 
shell-structured-inflatable froths. 
This classification, corresponds to the curvature of the manifold which the froth tiles. 
The space is elliptic for $|s|<2$, hyperbolic for $|s|>2$ and Euclidean for the fixed 
point $s=s^*=2$.
The map relate successive numbers $V^{(t)}$ of vertices per shell. 
Iterations of Eq.\TI\ generate trajectories in the plane $(t,V_+)$ starting from the 
initial points $V^{(-1)}_+=V_-^{(0)}=0$ and $V_+^{(0)}=\langle n \rangle$. 

When $|s| < 2$, the trajectories are given by the equation
\eqn\sst{  V^{(t)}_+=V^{(0)}_+ {\sin (\varphi (t+1) ) \over \sin \varphi} \;\ ,} %
with $\cos (\varphi) = {s/2}$.
Eq.\sst\ shows that all trajectories are finite and end at the point $V_+^{(T)} = 0$, 
with $T = {\pi \over \varphi}-1 $. Moreover, the values of $V_+^{(t)}$ are bounded 
by the quantity ${V_+^{(0)}/{\sin(\varphi)}}$.
These finite and bounded trajectories are describing the iterative tiling of compact 
manifolds with positive curvature. 
Indeed, consider a froth tiling the surface of a sphere.
Suppose that the north pole of the sphere is located in the germ cell; the successive 
shells are the parallels on the sphere.
The number of vertices per shell increases between the north pole and the equator, 
then decreases from the equator to the south pole where the tiling ends.
This is precisely the behaviour described by Eq.\sst. 
The quantity $T+1 = {{\pi}\over \varphi}$ is the topological distance between both 
poles. 
Here are a few examples of regular froths with $|s|<2$. To $s = -1$ corresponds to a 
froth made with four triangles, i.e. the surface of a tetrahedron. The recursion 
parameter $s = 0$ corresponds to a froth made with six squares, i.e. the surface of a 
cube. Finally, $s = 1$ is associated with a froth which is the surface of a 
dodecahedron.

In the case $|s|>2$, the solution of Eq.\TI\ is 
\eqn\ssht{  V^{(t)}_+=V^{(0)}_+ {\sinh (\varphi (t+1) ) \over \sinh \varphi} \;\;\;\; , 
}
with $\cosh (\varphi) = {s/2}$. Eq.\ssht\ shows that, contrary to the previous case, 
the values of $V_+^{(t)}$ increases exponentially with $t$. All trajectories are now 
infinite and unbounded in the plane ($t$,$V_+$). They are therefore describing the 
iterative tiling of non-compact manifolds with negative curvature. 

At the fixed point $s = s^{*} = 2$, Eq.\TI\ has the solution
\eqn\tt{ V_+^{(t)} = (t+1) V_+^{(0)} \;\;\;\; , }
The values of $V_+^{(t)}$ have again no upper bound, but here they are increasing 
linearly with $t$ as expected for the Euclidean plane by simple geometrical 
considerations. 
The fixed point $s^* = 2$ describes shell-structured-inflatable froths covering the 
Euclidean plane with cells with 6 edges on average. An  example of such froths is the 
hexagonal tiling. 

We have shown that the logistic map provides, in a natural way, the topological 
classification of tilings of manifolds  without any a-priori knowledge of their 
Gaussian curvature.
In 2D this classification by the logistic map is identical to that provided by the 
combination of the Gauss-Bonnet theorem \ref\KREYr\ and Euler's equation:
\eqn\gbe{\int \!\!\!\!\! \int {\kappa} \, d {a} = 
		{{\pi} \over 3}(6 - \langle n \rangle)F 
	= {{\pi} \over 3}(2 - s)F \;\;\;\; , }
here, $\kappa$ is the Gaussian curvature and it is integrated over the whole 
manifold. $F$ is the total number of cells in the manifold. 
The tiled manifold is hyperbolic, Euclidean or elliptic according when the integrated 
curvature is negative, zero or positive, i.e. when the recursion parameter $s$ is 
larger, equal to or smaller than 2.
However, the logistic map is also applicable in 3D where there is no Gauss-Bonnet 
theorem and the Euler equation is homogeneous \ref\AsRir .

\bigskip
\centerline{\bf  III. RECURSION PROCEDURE FOR 3D FROTHS}
\medskip
\nobreak
\xdef\secsym{3.}\global\meqno = 1
\medskip
This Section extends the analysis of the previous one to 3D shell-structured-
inflatable froths. 
The froth has  $V$ vertices, $E$ edges, $F$ faces and $C$ polyhedras. 
Every shell of the 3D froth is built up from two superposed different two-
dimensional froths, and looks like a corrugated sphere. 
This is the same as in 2D, where a shell can be regarded as the superposition of two 
1D froths, one whose vertices are connected to the ``incoming'' edges from shell $(t-
1)$ to shell $(t)$, and the other, whose vertices are connected to the ``outgoing'' 
vertices pointing from shell $(t)$ towards shell $(t+1)$. 
Similarly, every spherical shell $(t)$ of the 3D froth is built up of the superposition 
of two  2D froths, one whose edges are connected to the ``incoming'' faces of layer 
$(t-1)$, and the other whose edges are connected to the ``outgoing'' faces of layer 
$(t)$.
Let $V^{(t)}_{+(-)}$ and $E^{(t)}_{+(-)}$ be the numbers of vertices and edges of 
shell $(t)$, bounding the cells layer $(t)$ between shells $(t)$ and $(t+1)$ (layer $(t-
1)$ between shells $(t)$ and $(t-1)$, respectively), which are making the ``outgoing'' 
(``incoming'') froth. 
Let $F^{(t)}_{+(-)}$ be the number of faces of such froths.
Both froths are characterized by the identities
\eqn\FII{ V^{(t)}_{+(-)}- E^{(t)}_{+(-)} +F^{(t)}_{+(-)}= 2  \;\;\;\  , }
(Euler's formula) and
\eqn\FIII{3V^{(t)}_{+(-)} = 2 E^{(t)}_{+(-)} }
(since in both 2D froths, any vertex is connected by three edges and any edge is 
bounded by two vertices).

One has the following relations between two successive shells
\eqn\inca{\eqalign{V^{(t+1)}_- &= V^{(t)}_+    \cr E^{(t+1)}_- &= E^{(t)}_+      
\cr F^{(t+1)}_- &= F^{(t)}_+    \;\;\; .}}
Shell $(t)$ is a spherical surface tiled by a network with $F^{(t)}_N$ faces.
One has
\eqn\FI{F^{(t+1)}_N = \langle f \rangle F^{(t)}_+ - 2 E^{(t)}_+ - F^{(t)}_N \;\;\;\; 
.}
In this equation, $\langle f \rangle$ is the average number of faces per cell in the 
layer $(t)$.

Since the whole shell $(t)$ is a polyhedron, both Euler's formula and the incidence 
relations are applicable between the number of edges $E^{(t)}_{N}$, the number of 
vertices $V^{(t)}_{N}$ and the number of faces $F^{(t)}_{N}$ of the shell-
polyhedron, namely, 
\eqn\FIV{ V^{(t)}_{N}- E^{(t)}_{N} +F^{(t)}_{N}= 2 }
and
\eqn\FV{ \langle n \rangle_N F^{(t)}_N = 2E^{(t)}_{N} \;\;\;\; .}
Here, $\langle n \rangle_N$ is the average number of edges per face of the shell-
polyhedron. Since it is an elliptic tiling with vertex coordination $\ge 3$, then 
$\langle n \rangle_N < 6$. 
The shell-network is the superposition of two 2D froths, it has therefore 3-connected 
vertices $V_{+(-)}$ corresponding to the "outgoing" ("incoming") froth, and also 4-
connected vertices $V_{\times}$ at the intersections between edges of the two 2D 
froths. 
The three types of vertices are represented in Fig.(2). 
Fig.(3) shows the shell-network in the particular case of the ``Kelvin froth'' \ref\Kr, 
\ref\WILr, and indicates a 3-connected vertex and  a 4-connected vertex.

The total number of vertices $ V^{(t)}_{N}$ on shell $(t)$ is the sum of all 3- and 4-
connected  vertices, i.e.
\eqn\FVI{ V^{(t)}_{N}= V^{(t)}_{+} +V^{(t)}_{-}+V^{(t)}_{\times} \;\;\;\; .} 
The total number of edges $E^{(t)}_{N}$ on shell 
$(t)$ satisfies the equation
\eqn\FVII{2 E^{(t)}_{N}= 3V^{(t)}_{+} + 3V^{(t)}_{-}+ 
4V^{(t)}_{\times} \;\;\;\; .}
Using Equs.(3.5), (3.6) and (3.8), one obtains
\eqn\FVIII{ V^{(t)}_{N}= 2 + {1 \over 2} 
\Big (1- {2 \over \langle n \rangle_N}\Big )
	(3V^{(t)}_{+} + 3V^{(t)}_{-}+ 4V^{(t)}_{\times}) \;\;\;\; .} %
Combining Equs.(3.7) and (3.9), it is possible to express the variable 
$V^{(t)}_{\times}$ in terms of the variables  $V^{(t)}_{+}$ and $V^{(t)}_{-}$ 
alone as
\eqn\FIX{ 2V^{(t)}_{\times}={4 \langle n \rangle_N \over 4- \langle n \rangle_N} -
(V^{(t)}_{+} + V^{(t)}_{-})\Big ({6- \langle n \rangle_N \over 4- 
\langle n \rangle_N}\Big ) \;\;\;\; ,}
which,  with Eq.(3.6) and (3.8), yields
\eqn\FX{ F^{(t)}_{N} = 
{8- (V^{(t)}_{+} + V^{(t)}_{-}) \over 4- \langle n \rangle_N} 
\;\;\;\; .}
Putting Eq.(3.11) into Eq.(3.2), we obtain, with the help of Eqs.(3.1), (3.3) and (3.4), 
the following relation
\eqn\FXII{ V^{(t+1)}_{+} = {1 \over 2} 
\Big( (\langle f \rangle -6)(\langle n \rangle_N-4) - 
4 \Big) V^{(t)}_{+} - V^{(t)}_{-}+  
2\Big( 8+\langle f \rangle (\langle n \rangle_N-4) \Big) \;\;.  }
Finally, by shifting the variables $V_{+(-)}$ as 
\eqn\Shft{ \tilde{V}^{(t)}_{+(-)}= {V}^{(t)}_{+(-)}  -  
4{\Bigg ({8+\langle f \rangle (
\langle n \rangle_N-4) \over 8-(\langle f \rangle-6)(\langle n \rangle_N-4)} \Bigg) } 
\;\;\;\; ,}
one obtains the recursion equation
\eqn\Ffin{s \tilde{V}^{(t)}_+= \tilde{V}^{(t+1)}_+ + \tilde{V}^{(t-1)}_+ \;\;\;\; ,}
with the recursion parameter      
\eqn\S{s={1 \over 2} \Big( (\langle f \rangle-6)(\langle n \rangle_N-4)-4  \Big) \;\;\;\; 
.}
This recursion equation has the same matrix form as in the 2D case
\eqn\TII{ \Bigg(\eqalign{ &\tilde{V}^{(t+1)}_+ \cr &\tilde{V}^{(t+1)}_- } \Bigg)=
\Bigg(\eqalign{ &s \;\;\;\;  -1 \cr &1  \;\;\;\;\;\;\;\; 0} \Bigg) \Bigg(\eqalign{ 
&\tilde{V}^{(t)}_+ \cr &\tilde{V}^{(t)}_- } \Bigg) \;\;\;\; ,}
As in 2D $\langle f \rangle$ and $\langle n \rangle_N$ can be supposed to be 
independent of the distance $t$.
The variation of the 3D recursion parameter $s$ (the trace of the transfer matrix 
Eq.\TII\ ) is described by the logistic map (2.7), as in the 2D case. 
Consequently, the classification of 3D shell-structured-inflatable-froths is the same 
as in 2D. 

Elliptic shell-structured-inflatable froths are associated with $|s| < 2$. They are tiling 
iteratively 3D compact manifolds with positive curvature. 
Indeed, the corresponding solution of Eq.\TII\ is finite and bounded in the $(t,V)$ 
plane
\eqn\sstI{ V^{(t)}_+=A \sin (\varphi t +B) + 2 \Big ({8 + \langle f \rangle (\langle n 
\rangle_N-4) 
\over 2-s}\Big )\;\;\; , }
with $\cos (\varphi) = {s/2}$.
The coefficients $A$ and $B$ can be deduced from the initial conditions 
$V^{(0)}_+=2(\langle f \rangle-2)$ and $V^{(-1)}_+=0$.

Hyperbolic shell-structured-inflatable froths are associated with $|s| > 2$. They are 
tiling iteratively 3D non-compact manifolds with negative curvature. Indeed, the 
corresponding solution of Eq.\TII\ is unbounded in the $(t,V)$ plane
\eqn\sshtI{ V^{(t)}_+=A \sinh (\varphi t +B) + 
2 \Big ({8 + \langle f \rangle(\langle n \rangle_N-4) \over 2-s}\Big ) \;\;\; , }
with $\cosh (\varphi) = {s/2}$. 
As previously, the coefficients $A$ and $B$ can be determined from the initial 
conditions.

For $s=-2$ the solution reads
\eqn\smdue{ V^{(t)}_+=(-1)^t  \Big( A t + B \Big) + 
 {8 + \langle f \rangle (\langle n \rangle_N-4) \over 2}\;\;\; . }
with $A$ and $B$ deducible from the initial conditions.

The solution of Eq.\TII\ associated to the fixed point $s = s^* = 2$ is 
\eqn\ssdue{V^{(t)}_+ = (t+1) \Biggl(V_+^{(0)} + t \Bigl( 
	8+\langle f \rangle (\langle n \rangle_N -4 ) \Bigr) \Biggr) \;\;\;\; .} %
The quadratic dependence in $t$ is the one expected from simple geometrical 
reasoning for a tiling of the 3D Euclidean space. 

As in 2D, the logistic map gives a natural description of tilings of three dimensional 
manifolds without the need of any a-priori information on their curvature. 
Consequently, the logistic map is able to characterize curved manifolds even when 
the Gauss--Bonnet formula is not applicable \ref\POLr, \ref\CHEr. 
The generation of tilings of curved manifold by the recursion procedure has therefore 
a  wider applicability  than  the Gauss--Bonnet formula.

\bigskip
\centerline{\bf IV. EXAMPLES OF 3D SHELL-STRUCTURED} 
\centerline{\bf INFLATABLE FROTHS}
\medskip
\nobreak
\xdef\secsym{4.}\global\meqno=1
\medskip
In order to illustrate the previous considerations, we give some known examples of 
3D froths and show that they fit our classification. All are monotiled (i.e. constituted 
of topologically identical cells), apart from the last example.
 
The only regular elliptic froths in 3D are $\{3,3,3\}$ (packing of tetrahedra), 
$\{4,3,3\}$ (packing of cubes) and $\{5,3,3\}$ (packing of dodecahedra) \ref\COXr.
They correspond to $s = -1$, $s = -2$ and $s = 1$ respectively. 
Note that the case $s = 0$ does not correspond to any regular froth. 
Indeed, the only solution $s = 0$ of Eq.(3.15) with $\langle f \rangle$ and $\langle n 
\rangle_N < 6$ being both integers is $\langle f \rangle = 10$, $\langle n \rangle_N = 
5$; which is not regular.

Consider  Eq.(3.15) in the Euclidean case (i.e. $s=2$). This equation gives a 
relationship between the average number of neighbours per cell ($\langle f \rangle$) 
in the 3D froth and the average number of edges per cell ($\langle n \rangle_N$) in 
the 2D spherical shell-network
\eqn\due{\langle f \rangle= 6 + {8 \over \langle n \rangle_N -4 } \;\;\;\; . }
This equation gives the condition for Euclidean space-filling by a shell-structured-
inflatable froth. 
Note that, from Eq.(4.1), the minimal number of faces per cell of such a froth is 10, 
since $\langle n \rangle_N < 6$.
It is known that the minimal number of neighbours per cell is 8 for an Euclidean 
froth. Thus an Euclidean froth with $8 \le \langle n \rangle_N < 10$ necessarily 
contains local topological defects of the kind shown in Fig.(11).

Recall that the shell-network is the superposition of two elliptic 2D froths, the 
``incoming'' and the ``outgoing'' froths. 
The pattern of edges constituting the shell-network sets the value of $\langle n 
\rangle_N$. 
Therefore, Eq.(4.1) allows us to construct systematically 3D Euclidean shell-
structured-inflatable froths starting from the 2D shell-networks. 

The simplest 2D froth is the hexagonal lattice. 
The examples displayed in Figs.(4), (5) and (6) illustrate the construction of ordered, 
monotiled 3D froths from a shell-network generated by different superpositions (cf. 
Figs.(4a), (5a) and (6a)) of two hexagonal lattices. 
Fig.(4b, c) shows two 3D unit cells constructed from the network (4a) (see also 
\ref\GLAZr ). The cell (4b) is topologically equivalent to Kelvin's $\alpha$--
tetrakaidecahedron \Kr\  \WILr  (it builds up the Kelvin froth shown in Fig.(3)), and 
the cell (4c), to its twisted variant \ref\Wr . 
Both structures have $\langle f \rangle = 14$ and $\langle n \rangle_N = 5$, 
Eq.(3.15) gives $s=2$. They are indeed Euclidean space-fillers.

Fig.(5a) shows part of a shell-network with 5-sided faces, generated by the 
superposition of two ``squeezed'' hexagonal lattices (see also \GLAZr ). 
Figs.(5b) and (5c) show two 3D unit cells constructed from the network (5a). 
These cells have again $\langle f \rangle = 14$. 
The unit cell (5b) is topologically equivalent to the $\beta$--tetrakaidecahedron (the 
Williams cell \ref\WILSr ). 
It has $\langle n \rangle_N = 5$, and is an Euclidean space-filler according to 
Eq.(3.15).

The unit cell of Fig.(5c) is topologically equivalent to the 14-sided cell (the 
Goldberg cell \ref\WEGr ) which occurs, among others, in clathrates \WILr , in t.c.p. 
structures \ref\FRKr \ref\SHOr\ and in the minimal froth of Weaire and Phelan 
\ref\WPr. 
Space can be filled layer by layers of Goldberg cells only. 
The layers (Fig.(5) ) are Euclidean and the network (5a) is the same as that of the 
Williams space-filler.
However, successive layers are more and more distorted  \ref\WEDISr , as shown in 
Fig.(5d).
This distortion, which stretches the network in one direction and compresses it in the 
other, strongly suggests that we are filling  hyperbolic 3D space with a stack of 
Euclidean layers.
It is possible to prove this contention by filling space shell by shell instead of layer 
by layer. 
When doing so, one finds that most of the shell-network is composed of pentagons 
(12 out of 14 in each 3D cell), but a finite density of hexagons (2 out of 14 in each 
3D cell) is needed in order to close a shell. 
Thus, $\langle n \rangle_N > 5$ which, according to Eq.(3.15), implies $s > 2$. 
Hence, the 3D manifold tiled by Goldberg cells is hyperbolic. 

With another intersection of the two ``squeezed'' hexagonal lattices, one generates 
the shell-network shown in Fig.(6a). 
The corresponding 3D unit cell (6b), has $\langle f \rangle = 16$ (8 quadrilaterals, 6 
hexagons and 2 octagons) and $\langle n \rangle_N = 4.8$. As far as we know, this 
unit cell is a new monotile Euclidean space-filler.

Fig.(7) shows an example of an Euclidean shell-structured-inflatable froth made of 
two different cells \WILr. The shell-network (7a) has also two different tiles. The 
associated 3D unit cell (7b) has $\langle f \rangle = 12$.

One can  show in general that any Euclidean shell-structured-inflatable froth made 
with topologically identical cells can be constructed from a shell-network generated 
by superposition of two hexagonal lattices. 

The construction of 3D disordered froths from 2D disordered shell-networks is 
discussed in Appendix C.

Although a construction of 3D froths layer by layer has been given in \GLAZr , it 
must be emphasized that our approach, combining spherical shells with the logistic 
map, is  more general and provides a unifying way to deal with 3D space-filling 
structures, whether regular or not, whatever the curvature of the manifold which they 
are tiling.

\bigskip
\centerline{\bf V.  BOUNDS ON TOPOLOGICAL PROPERTIES OF NATURAL} 
\centerline{\bf CELLULAR SYSTEMS AND T.C.P. STRUCTURES}
\medskip
\nobreak
\xdef\secsym{5.}\global\meqno=1
\medskip
The average number ($\langle n \rangle$) of edges per face of a 3D froth is in 
general different from the average number of edges per face in the shell-network 
($\langle n \rangle_N$).
For example, the froths in Figs.(4) and (5b) have $\langle n \rangle_N = 5 $ and 
$\langle n \rangle = 5.14 $, the froth in Fig.(6) has $\langle n \rangle_N = 4.8 $ and 
$\langle n \rangle = 5.25 $ and the froth in Fig.(7) has $\langle n \rangle_N = 5.33 $ 
and $\langle n \rangle = 5 $.

The value of $\langle n \rangle$ is related to the average number of faces per 3D cell 
by %
\eqn\quattro{\langle f \rangle={12 \over  6-\langle n \rangle} \;\;\;\; . } 
It is interesting to study the competition between Eq.(5.1) and the Euclidean space-
filling condition given by Eq.(4.1). 
These two relations $\langle f \rangle (\langle n \rangle_N)$  (labelled ``space-
filling'') and $\langle f \rangle (\langle n \rangle)$ (labelled ``3D cell'') are plotted in 
Fig.(8). 
They meet at the point ($\langle n \rangle^* , \langle f \rangle^*$) given by 
\eqn\cinq{ \eqalign{\langle n \rangle^* &= {10 + 2\sqrt 7 \over 3} \cr \langle f 
\rangle^*&=8 + 2\sqrt 7  } \;\;\;\; .}
It is only when the equality $\langle n \rangle=\langle n \rangle_N= \langle n 
\rangle^*$ (which corresponds to $\langle f \rangle^* =  13.29...$) is satisfied that an 
arbitrary cell has the freedom to adhere to a preexisting shell by any subset of it 
faces, without adjustment. 
This freedom grants therefore a larger number of possibilities for building up a froth 
and it maximizes the orientational entropy per cell. 
Indeed, Eq.(5.1) is a constraint on any single 3D cell, whereas Eq.(4.1) is a constraint 
on the set of 3D cells in a layer. 
When $\langle n \rangle=\langle n \rangle_N=\langle n \rangle^*$, one of the two 
constraints is automatically satisfied by the other and the orientational entropy is 
increased \ref\RILIr.
 
Note that the value $\langle f \rangle^* = 13.29...$  falls within the range of several 
already known bounds.
It is consistent with the values $13.2$ and $13.33...$ resulting from the decurving of 
the dodecahedral packing with 14- and 18-sided cells or 14- and 16-sided cells, 
respectively \ref\SMr.
Kusner \ref\KUr\ has shown that a single cell with minimal interfaces in a froth 
which is locally Euclidean or hyperbolic cannot have less than $13.39...$ faces on 
average. 
It is also known that the minimal number of faces per cell of a periodic, monotiled 
froth is 14.
Weaire and Phelan have recently given an example of froth with $\langle f 
\rangle=13.5$ (the so-called A15 phase) which minimizes the total interfacial area 
\WPr (see also \ref\RIVr ). 

Natural froths minimize their free energy ({\it Configurational Energy} minus {\it 
Temperature $\times$ Entropy}).
With the bounds given above, this condition is realized when 
the value of $\langle f \rangle$ is between 13.29... and 13.5 (or 14 for periodic 
monotiled froths). 
The lower bound corresponds to configurations with maximal orientational entropy, 
whereas the upper bound corresponds to configurations with minimal interfacial 
energy.

There exists a class of natural structures, the Frank and Kasper phases (or the larger 
class of t.c.p. structures \FRKr, \SHOr), for which $\langle f \rangle$ falls within 
these bounds. 
These structures are periodic and made of 12-, 14-, 15- and 16-sided cells whose 
faces are either pentagons or hexagons. 
It can be verified that some of them fulfil the condition of Euclidean space-filling 
given by Eq.(4.1). 
We can therefore assume that the t.c.p. structures are Euclidean shell-structured-
inflatable froths. 
Then, their shell-network is a periodic tiling made of pentagons and hexagons only. 
Let the 2D unit cell of the shell-network consist of $f^{(5)}$ pentagons and 
$f^{(6)}$ hexagons, belonging to $N^* $ polyhedra within the layer between two 
subsequent shells.
The number of polyhedra in the 3D unit cell is a multiple of $N^* $.
The average number of edges per face in the shell-network is 
\eqn\ffkI{ \langle n \rangle_N = { 6  f^{(6)} + 5  f^{(5)} \over f^{(6)} + f^{(5)} } 
\;\;\;\; . }
Substituting  into Eq.(4.1), one obtains 
\eqn\ffkII{ \langle f \rangle ={ 20  f^{(6)} + 14  f^{(5)} \over
	2 f^{(6)} + f^{(5)} } \;\;\;\; . }
The number of polyhedra in the 3D unit cell can be calculated with the help of the 
numbers of faces the ``outgoing'' ($f_+$) and ``incoming'' ($f_-$) froths in the unit 
cell of the shell-network. 
These numbers coincide with the numbers of polyhedra in the layers above $(f_+)$ 
and below $(f_-)$ the shell which have one or more faces belonging to the 2D unit 
cell.
In the limit of large shell-networks, one has the relation  $v_{+(-)} \simeq 2f_{+(-
)}$, with  $v_+$ (resp. $v_-$) counting the number of 3-connected vertices in the 2D 
unit cell which belong to the ``outgoing'' (resp. ``incoming'') froths.
Eq.\FIX\ can then be written in terms of the quantities associated with the 2D unit 
cell only
\eqn\ffkIII{ f_+ + f_- = v_\times \Big({ \langle n \rangle_N -4
		\over 6- \langle n \rangle_N }\Big ) } %
($v_\times$ counts the number of 4-connected vertices in the 2D unit cell).
On the other hand, since the 3D system is periodic, one has $f_+ + f_- = 2 N^*$ on 
average. 
Therefore Eq.(5.5) can be written as
\eqn\fkIV{ N^* = {v_\times \over 2} \Biggl( 1 + 2
{ f^{(6)} \over f^{(5)} } \Biggr) \;\;\;\;.}
If one puts into Eq.(5.4) the simplest  combinations of integers $f^{(5)}$ and 
$f^{(6)}$ which are such that $\langle f \rangle$ falls within the two bounds
$13.29...$ and $13.5$, one retrieves the average number of faces of the 3D unit cell 
of all experimentally known t.c.p., which are  listed in Table 1.
(The Table gives all the possible combinations $(f^{(5)} , f^{(6)})$ up to $f^{(6)} = 
4$ and, for $f^{(6)} \ge 4$, only those corresponding to known natural structures.) 
Also given are the corresponding values of $N^*$, obtained from Eq.(5.6). 
These  values of   $N^*$ are exactly equal to the sum of the lowest non congruent 
numbers of 16- ($p$), 15- ($q$), 14- ($r$) and 12-sided polyhedra ($x$) in the 
structural formula of the corresponding  t.c.p. \SHOr. 
The Table presents also several simple combinations ($f^{(6)} , f^{(5)}$) which 
correspond to structures not (yet) observed (they are indicated by blanks in the last 
column). 
Notably, combinations (2,23), (2,25)... may be good candidates for t.c.p. structures 
yet to be observed. 
On the other hand, combinations (2,19), (3,28), (3,29) and (4,39) may be too 
distorted to qualify as t.c.p. structures. They may be realized with atoms of very 
different sizes.
Note finally, that when $\langle f \rangle$ is represented as a function of the ratio 
$f^{(5)}/f^{(6)}$, the structures in the Table 1 tend to gather into distinct groups. 
This may indicate either the existence of unfavourable configurations or  structural 
mode-locking into the simplest t.c.p. structures (A15, Z, $\sigma$, ... C15).

All these facts strongly suggests that the t.c.p. are shell-structured-inflatable froths.  

\bigskip
\centerline{\bf  VI. CONCLUSION}
\medskip
\nobreak
\xdef\secsym{5.}\global\meqno = 1
\medskip
In this paper we have introduced a new way to study froths, which emphasizes their 
shell structure. We have studied an important subclass of shell--structured froths, i.e. 
those which can be generated in a recursive way according to an inflationary 
procedure.
For 2D froths (and networks with any coordination number) and 3D froths we have 
found that this recursive procedure is described by the logistic map. This map allows  
for a natural differentiation between froths tiling elliptic, hyperbolic or Euclidean 
manifolds, without any a--priori imposed curvature condition.
In particular, the logistic map is able to characterize 3D curved manifolds, thereby 
providing a way to define the curvature from topological considerations when the 
Gauss--Bonnet theorem is not applicable.
The logistic map in 3D enables us recover known space-filling configurations, and 
also to suggest new ones. 
It is clear that the approach using the logistic map is very powerful, since 
classification of 3D space--filling configurations is reduced to the study of the 2D 
tiling of the (elliptic) shell--surface.  
As an example of the power and generality of this approach, we have been able to 
retrieve the topological properties of all experimentally-known t.c.p. structures by 
studying the tiling of the shell--surface by pentagons and hexagons.

\bigskip
\centerline{\bf ACKNOWLEDGEMENTS}
\nobreak
\medskip
\nobreak
We are grateful to D. Weaire for many discussions, and to the referee for making us 
strengthen our conclusion. 
This work has been supported in part by the E.U. Human Capital and Mobility 
Program "Physics of Foams", ref. CHRX-CT-940542.
\nobreak
\par 
\bigskip
\vfill
\eject
\goodbreak
\centerline{\bf APPENDICES}
\xdef\secsym{A.}\global\meqno = 1
\bigskip
\nobreak
{\bf A  INFLATION OF TWO DIMENSIONAL  Z-VALENT NETWORKS WITH 
Z $\ge$ 4}

The generalization of Eq.\TI\ to the description of 2D shell-structured-inflatable 
froths with coordination number $z \ge 4$ is as follows. 
Every shell has $(z-1)$ different types of vertices. Extending the notation of Section 
2, the various types of vertices are labelled by $V^{(t)}_a$ with $a = 0,1,...,z-2$, 
$V_a^{(t)}$ being the number of vertices belonging to shell $(t)$ from which $a$ 
(resp. $z-2-a$) edges are pointing towards shell $(t+1)$ (resp. shell $(t-1)$). 
Every vertex $V_{a}^{(t)}$ adds $a$ cells between shells $(t)$ and $(t+1)$. 
The total number of cells $F^{(t)}$ between the two shells is 
\eqn\totN{ F^{(t)}= \sum_{a=0}^{z-2} a V_a^{(t)} = \sum_{a=0}^{z-
2} (z-a-2) V_a^{(t+1)}\;\;\; . }
Let $\langle n \rangle$ denote the average number of edges per cell layer $(t)$. 
If one sums over all cells in this layer, one obtains 
\eqn\totNI{\langle n \rangle F^{(t)} = \sum_{a=0}^{z-2} (a+1) V_a^{(t)} + 
\sum_{a=0}^{z-2} (z-a-1) V_a^{(t+1)}\;\;\; .  }
Since
\eqn\ident{a+1 = \Big(1+ {1 \over z-2} \Big)a + \Big({1 \over z-2} \Big) (z-a-2)} 
and
\eqn\identry{z-a-1 = \Big(1+ {1 \over z-2} \Big)(z-a-2) + \Big({1 \over z-2} \Big)a  
\;\;\;\  , }
one has
\eqn\totNI{\eqalign{
\langle n \rangle F^{(t)} &=  
\Big( 1+ {1 \over z-2} \Big) \sum_{a=0}^{z-2} a V_a^{(t)} + 
\Big({1 \over z-2} \Big) \sum_{a=0}^{z-2} (z-a-2) V_a^{(t)}\cr
&+ \Big({1 \over z-2} \Big) \sum_{a=0}^{z-2} a V_a^{(t+1)} + \Big(1+ {1 \over z-
2} \Big) \sum_{a=0}^{z-2} (z-a-2) V_a^{(t+1)}\cr &= \Big( 1+ {1 \over z-2} \Big) 
F^{(t)} + \Big({1 \over z-2} \Big) F^{(t-1)}\cr
&+ \Big( {1 \over z-2} \Big) F^{(t+1)} + \Big(1+{1 \over z-2} \Big) F^{(t)}
\;\;\;\; . } }
One obtains finally the recursion relation
\eqn\totNII{ \Big( \langle n \rangle(z-2)-2(z-1) \Big) F^{(t)}=F^{(t+1)}+F^{(t-1)} 
\;\;\;\; . }  
The matrix form of this recursion relation is the same as for $z=3$ (Eq.(2.3) ), with 
recursion parameter  $s=\langle n \rangle(z-2) - 2(z-1)$. 
The initial conditions are $F^{(1)}=(z-2) \langle n \rangle$ and $F^{(0)} = 1$.

Euclidean tilings are associated with the fixed point $s^{*} = 2$, i.e. to the equation
\eqn\nZ{ \langle n \rangle = {2z \over z-2} \;\;\; .}
The only regular solutions ($z$ and $\langle n \rangle$ integers) of this equation are 
$(6,3)$ (tiling by triangles), $(4,4)$ (tiling by squares) and $(3,6)$ (tiling by 
hexagons) (dual of $(6,3)$). 

\bigskip
{\bf   B.1  SHELL-STRUCTURED BUT NON-INFLATABLE 2D FROTHS} 
\xdef\secsym{B.1.}\global\meqno = 1
\nobreak
\medskip
\nobreak
Some 2D shell-structured froths cannot be constructed according to the recursion 
procedure of Eq.(2.3). 
These froths have local inclusions which are topological defects in the recursion 
procedure. 
An inclusion in a layer is a cell with neighbouring cells in this layer and only in one 
of the two neighbouring layers. 
Topological defects fall in two classes: vertex decorations (Figs.(9a) and (9b)) and 
edge decorations  (Fig.(9c)). 
In all cases the inclusion is on the $+$ side of shell $(t)$.

Defects can be eliminated by removing one or more edges and its surrounding 
vertices.
A  vertex-decoration defect is then replaced by an ordinary vertex (Fig.(10a)). An 
edge-decoration defect is then replaced by edges on the shell (Fig.(10b)).

The removal of one edge reduces by one unit the number of faces in the layer. 
This operation corresponds to the transformations $E \rightarrow E-3$, $V 
\rightarrow V-2$, $F \rightarrow F-1$. 
Consequently, since $\langle n \rangle = {2E \over F}$, the average number of edges 
per cell changes as 
\eqn\one{ \langle n \rangle'=\langle n \rangle+{\displaystyle 1 \over F-1} (\langle n 
\rangle-6) \;\;\;\; .}
The recursion parameter $s=\langle n \rangle-4$ changes therefore as
\eqn\two{s'=s+{\displaystyle 1 \over F-1} (s-2) \;\;\;\;. }
One sees that the fixed point $s^{*} = 2$ remains unchanged by defect elimination. 
Moreover, elliptic froths become more elliptic (i.e. $\langle n \rangle' < \langle n 
\rangle  < 6$) whereas hyperbolic froths become more hyperbolic (i.e. $\langle n 
\rangle' > \langle n \rangle  > 6$). 
Thus, the Euclidean, hyperbolic or elliptic character of the manifold tiled by the froth 
is not modified by defect elimination (it is indeed given by the Euler-Poincar\'e  
characteristic which is a topological invariant).

\bigskip
{\bf    B.2   SHELL-STRUCTURED BUT NON-INFLATABLE 3D FROTHS}
\nobreak
\medskip
\nobreak
\xdef\secsym{B.2.}\global\meqno = 1
\nobreak
By analogy with the 2D case, one can define a topological
distance $r$ between two cells
$A$ and $B$ as the minimal number of faces that must be crossed by a
path that connects $A$ and $B$. A 3D
shell-structured-inflatable froth is defined by the following two conditions :

\noindent
1) For any set of cells equidistant from a germ cell, there exists a closed non self-
intersecting surface that cuts these cells and no others.

\noindent
2) Any cell at distance $t$ from the germ cell is the neighbour of at least one cell at 
distance $t+1$.

\noindent
Shells are closed surfaces tiled by faces of cells; they bound layers of equidistant 
cells. It is possible to connect two adjacent shells $(t)$ and $(t+1)$ through a set of 
faces, each with one edge on shell $(t)$ and one on shell $(t+1)$.
Shell $(t)$ separates the whole froth into an inner froth, constituted of cells at 
distances $r \le t$, and an outer froth, with cells at distances $r>t$.

There are local defects which violate rules (1) or (2). 
These non-inflatable configurations in 3D froths are shown in Fig.(11). These are 
particular examples of the three general classes of 3D topological defects: vertex, 
edge and face decoration. as in 2D these non-inflatable configuration can be 
eliminated.
Defects elimination is made by removing one (or more) face(s), together with 
surrounding edges and vertices.
The removal of one face with $n$ edges reduces by one unit the total number $C$ of 
cells. 
This operation corresponds to the transformation $C \rightarrow C-1$ and 
$F \rightarrow F-1-n$. 
Consequently, since $\langle f \rangle = {2F \over C}$, the average number of faces 
per cell changes as 
\eqn\three{\langle f \rangle'=\langle f \rangle+{\displaystyle 1 \over C-1} 
\Bigl(\langle f \rangle-2(n+1) \Bigr)\; . }
In contrast to the 2D case, this transformation depends on the parameter $n$. 
This is not surprising since it is well-known that in 3D the value of $\langle f 
\rangle$ is not directly related to the curvature of the manifold tiled by the froth.

\bigskip
\centerline{\bf   C  RANDOM 3D EUCLIDEAN FROTHS FROM }
\centerline{ \bf  2D RANDOM SHELL NETWORKS} 
\xdef\secsym{C.}\global\meqno = 1
\medskip
\nobreak
\medskip
Eq.\due\ implies that a 3D random froth can be constructed from the superposition 
of two 2D random froths. To study this general case it is useful to rewrite Eq.\due\ in 
term of the number $p^\times$ of intersections of edges of the incoming froths by 
edges of the outgoing froths and vice versa.
For a given shell $t$ this quantity is equal to
\eqn\pX{p^\times={2V_\times^{(t)} \over E_+^{(t)} + E_-^{(t)}}=
                 {2 \over 3} {2 V_\times^{(t)} \over (V_+^{(t)} + V_-^{(t)})}
 \;\;\;\; ,}
where we used the identity $3V^{(t)}_{+(-)}=2E^{(t)}_{+(-)}$.
Using equation \FIX\ it is possible to express $p^\times$ in terms of 
$\langle n \rangle_N$. One has
\eqn\pXI{ p^\times = {2 \over 3} \Big( {6-\langle n \rangle_N \over 
\langle n \rangle_N-4} - {4\langle n \rangle_N \over (\langle n \rangle_N - 4)
(V_+^t+V_-^t)} \Big) \;\;\;\; .}
When the number of network cells is much larger than unity, one has $ p^\times 
=(2/3)(6-\langle n \rangle_N)/(\langle n \rangle_N-4)$. 
Substituting into \due, one obtains
\eqn\tre{ \langle f \rangle = 10 + 6 p^\times \;\;\;\; .}

In principle, in random froths, $p^\times$ can take any value between zero and
 infinity (but only between $2/3$ and 1 for periodic monotiled froths). For 
example, $p^\times= \infty$ corresponds to a froth made with layers of 
infinitely long bricks disposed, layer by layer, with orientation alternating 
by $90^o$. In this case the network is a square lattice.
The opposite limit ($p^\times =0$) corresponds, for example, to a 3D froth 
made with layers of large and small cells, when the ratio between the cell--
sizes tends to infinity. In this case the network is the result of the 
superposition of a froth with cells of large sizes and a froth of small sizes 
and the probability of intersection of two edges of these two froths tends 
to zero.

\vskip1.cm
\bigskip
\centerline{\bf REFERENCES}
\medskip
\nobreak
\item{\WRr}\WR
\item{\STAr}\STA
\item{\SCHUr}\SCHU 
\item{\LAVr}\LAV
\item{\KREYr}\KREY
\item{\AsRir}\AsRi
\item{\Kr}\K
\item{\WILr}\WIL
\item{\POLr}\POL 
\item{\CHEr}\CHE 
\item{\COXr}\COX
\item{\GLAZr}\GLAZ
\item{\Wr}\W
\item{\WILSr}\WILS 
\item{\WEGr}\WEG
\item{\FRKr}\FRK
\item{\SHOr}\SHO
\item{\WPr}\WP
\item{\WEDISr}\WEDIS
\item{\RILIr}\RILI
\item{\SMr}\SM
\item{\KUr}\KU
\item{\RIVr}\RIV
\bigskip
\vfill
\goodbreak
{\bf FIGURE CAPTIONS}
\vskip.7cm

\noindent
Fig.1 
\nobreak
\medskip
\nobreak
Schematic picture of a 2D shell-structured-inflatable froth.  

\vskip.7cm
\noindent
Fig.2 
\nobreak
\medskip
\nobreak
Any 3D shell $(t)$ is tiled with a network generated by the intersection of the faces 
coming to and going away from its surface. 
This shell-network has 4-connected vertices $V_{\times}^{(t)}$ (whose all 4 edges 
are belonging to the shell-network) and 3-connected vertices $V_{+(-)}^{(t)}$ (with 
3 edges belonging to the shell-network and the last one going away from it). 

\vskip.7cm
\noindent
Fig.3 
\nobreak
\medskip
\nobreak
An example of 3D shell-structured-inflatable froth, the Kelvin froth. A portion of the 
shell-network is brought out by hatcheries.

\vskip.7cm
\noindent
Fig.4 
\nobreak
\medskip
\nobreak
The two 3D space-filling unit cells constructed from the shell-network (a). The cell 
(b) is topologically equivalent to Kelvin's $\alpha$--tetrakaidecahedron and the cell 
(c) to its twisted variant. Both have 14 faces.

\vskip.7cm
\noindent
Fig.5 
\nobreak
\medskip
\nobreak
The two 3D space-filling unit cells constructed from the  shell-network (a) generated 
by the superposition of two ``squeezed'' hexagonal lattices. 
The cell (b) is topologically equivalent to the $\beta$--tetrakaidecahedron. 
The cell (c) is topologically equivalent to the Goldberg cell.

\vskip.7cm
\noindent
Fig.6 
\nobreak
\medskip
\nobreak
The 3D space-filling unit cell (b) (which has 16 faces) resulting from the shell-
network (a) generated by the superposition of two ``squeezed'' hexagonal lattices. 

\vskip.7cm
\noindent
Fig.7 
\nobreak
\medskip
\nobreak
Example of a 3D periodic shell-structured-inflatable froth (with $\langle f \rangle 
=12$) whose unit cell has two different elementary cells. 
(a) Shell-network. (b) 3D unit cell. 

\vskip.7cm
\noindent
Fig.8  
\nobreak
\medskip
\nobreak
The average number $\langle f \rangle$ of faces per cell in a froth plotted as a 
function of the average number $\langle n \rangle_N$ of edges per 2D cell in the 
shell network (Eq.(4.1), curve labelled "space-filling") and of the average number 
$\langle n \rangle$ of edges per face in the froth (Eq.(5.1), curve labelled "3D"). The 
abscissa $n$ represents both $\langle n \rangle_N$ and $\langle n \rangle$. 

\vskip.7cm
\noindent
Fig.9
\nobreak
\medskip
\nobreak
Local topological defects in the 2D recursion procedure. Figures (a) and (b) are 
examples of vertex decorations whereas Figure (c) is an example of edge decoration. 
The index $t$ denotes the topological distance.  

\vskip.7cm
\noindent
Fig.10
\nobreak
\medskip
\nobreak
Schematic representations of the elimination of a 2D local topological defect. (a) 
Vertex decoration. (b) Edge decoration.

\vskip.7cm
\noindent
Fig.11
\nobreak
\medskip
\nobreak
Local topological defects in the 3D recursion procedure, (a) is a vertex decoration 
defect, (b) is an edge decoration defect and (c) is a face decoration defect. The index 
$t$ denotes the topological distance.

\vskip.7cm
\bigskip
{\bf TABLE CAPTION}
\nobreak
\medskip
\nobreak
\noindent
Table 1 
\nobreak
\medskip
\nobreak
Average number of faces $\langle f \rangle$ and (minimal)  number of elements in 
the 3D unit cell $N^*$ of all the t.c.p. structures known experimentally (labelled in 
the last column) and of hypothetical t.c.p. structures (indicated by a blank in the last 
column). The integers $p$, $q$, $r$ and $x$ indicate respectively the proportions of 
3D cells with 16, 15, 14 and 12 faces present in the 3D unit cell.

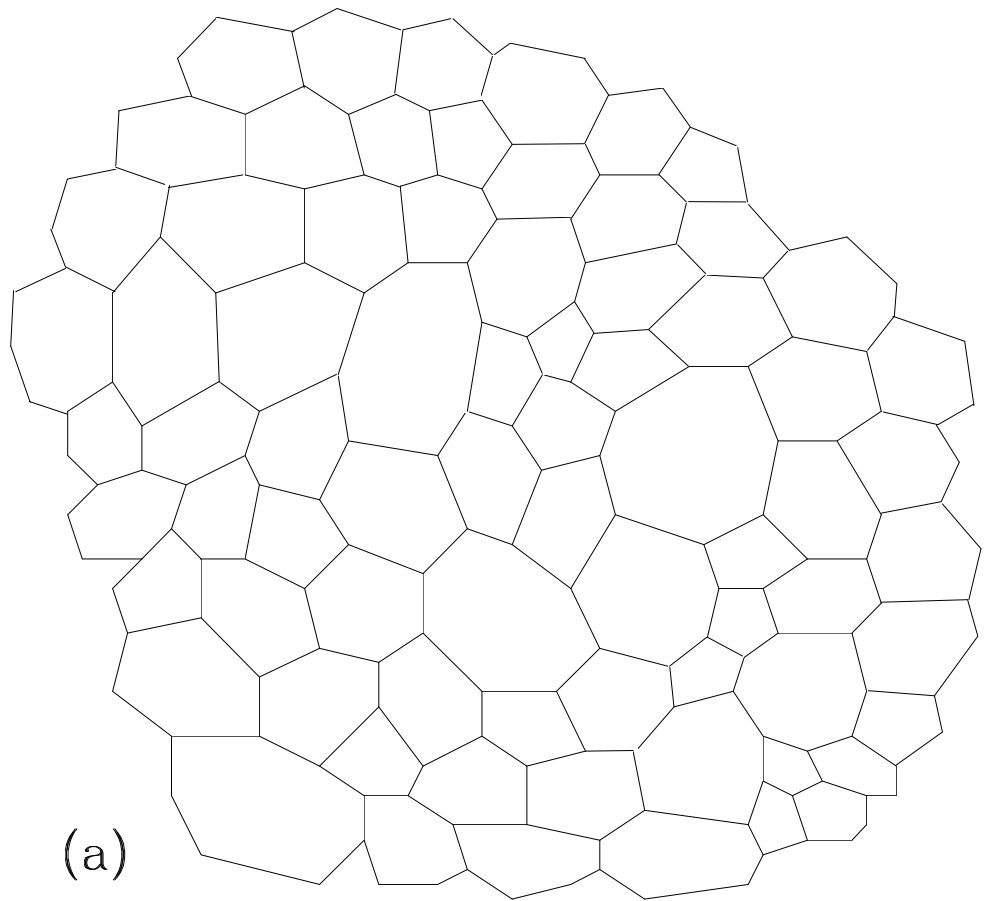
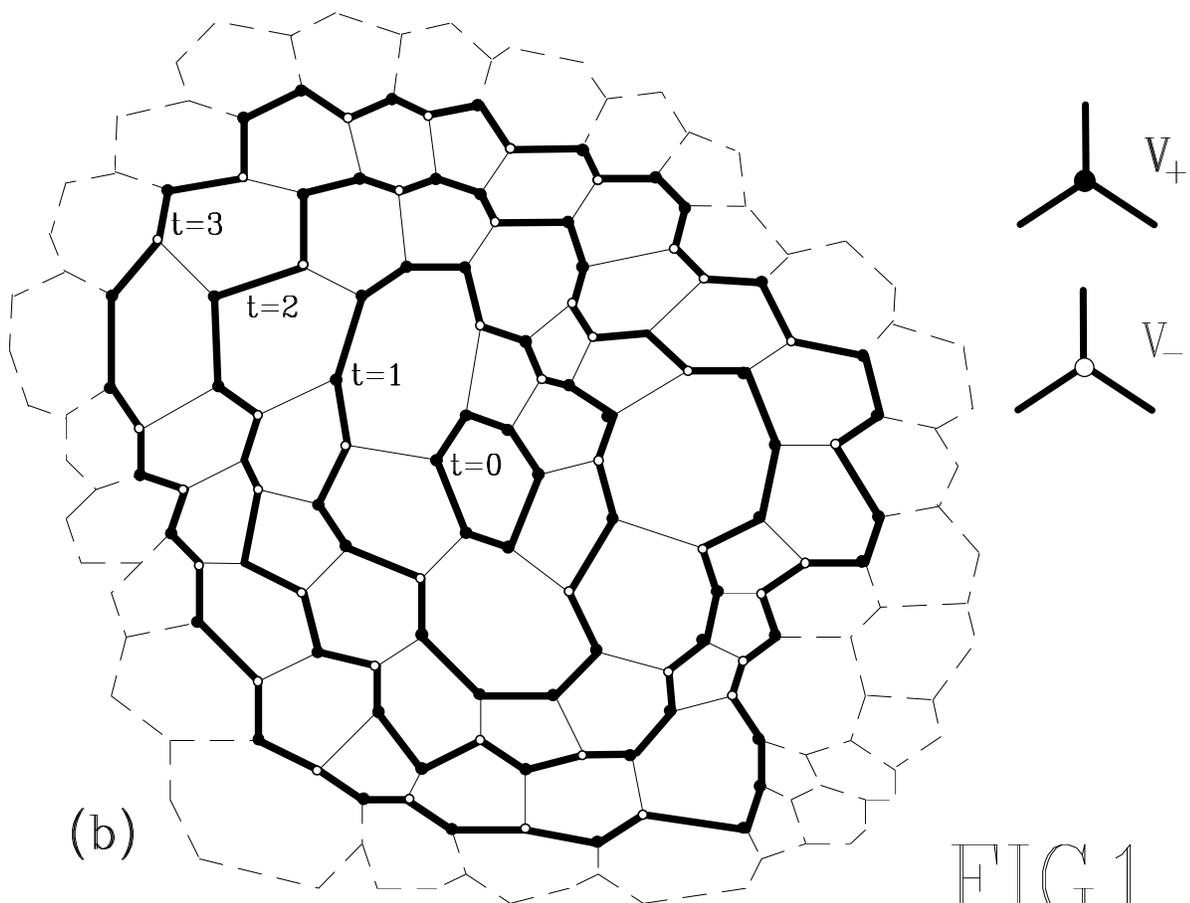
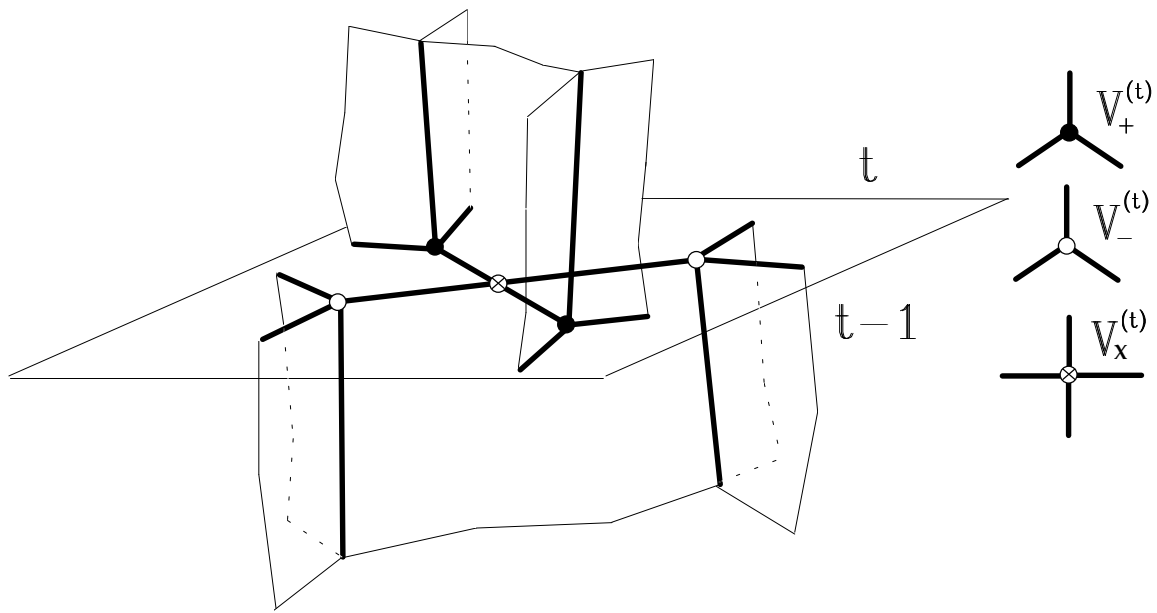
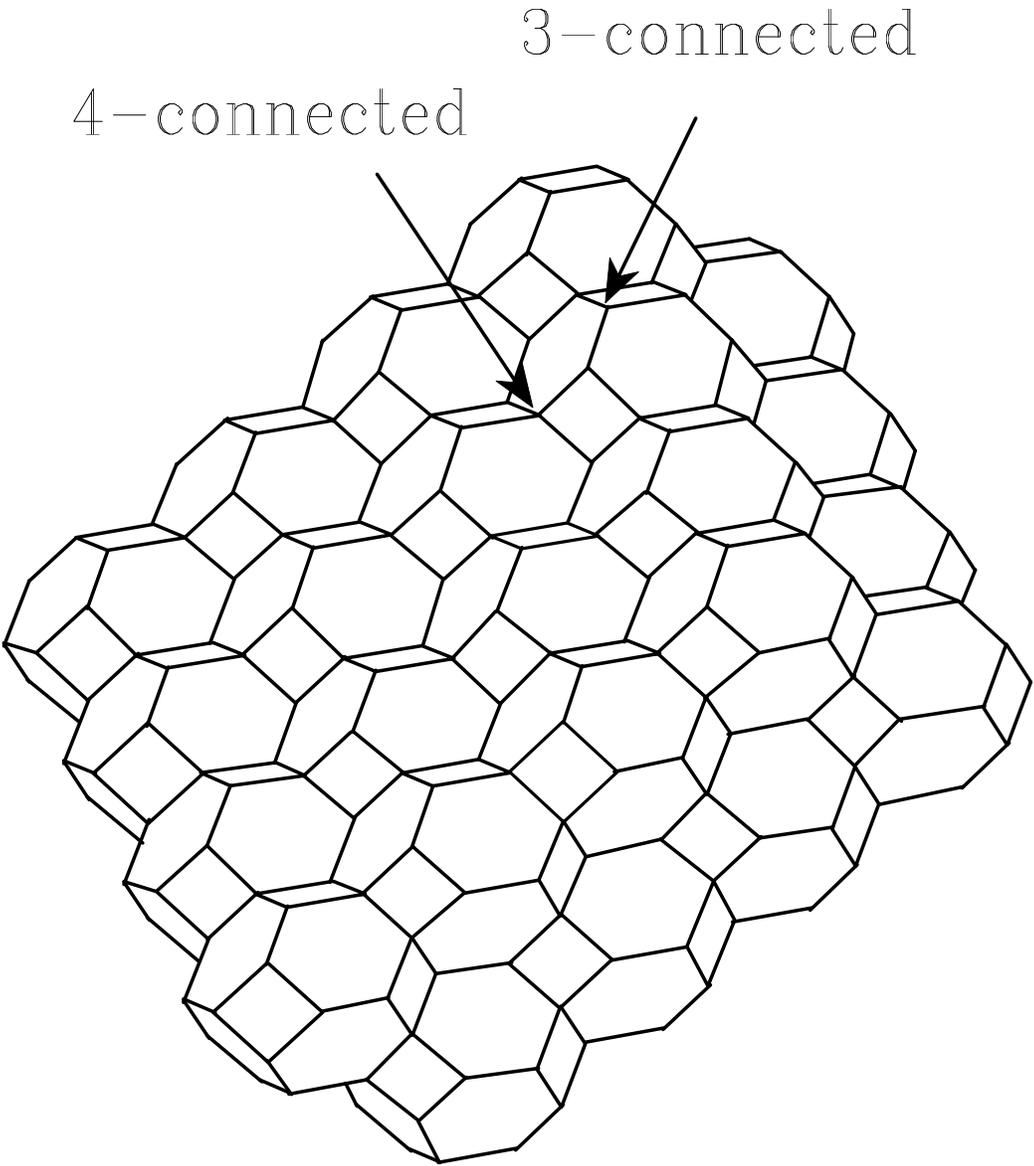
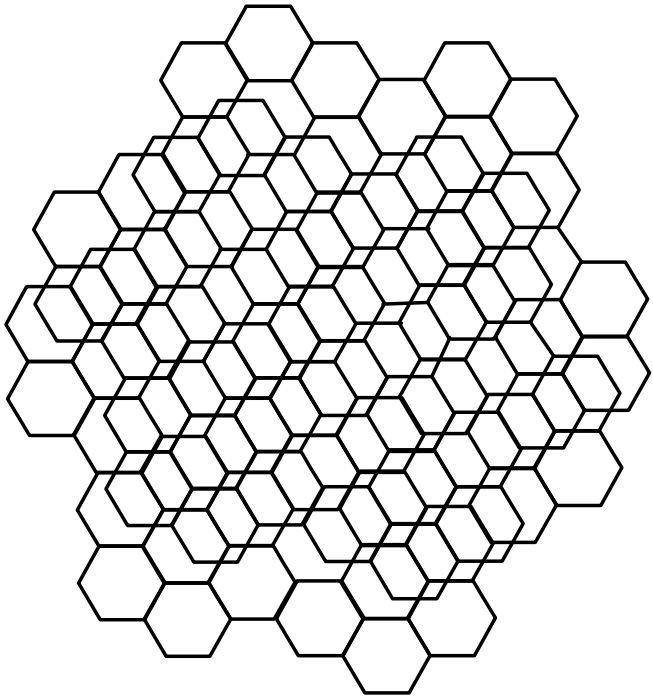
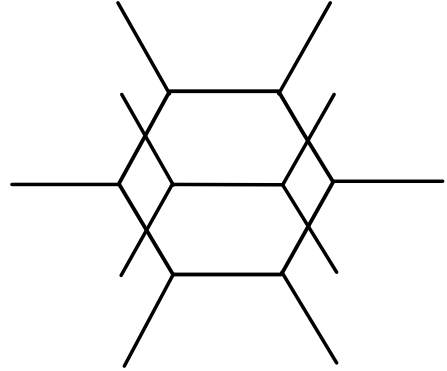
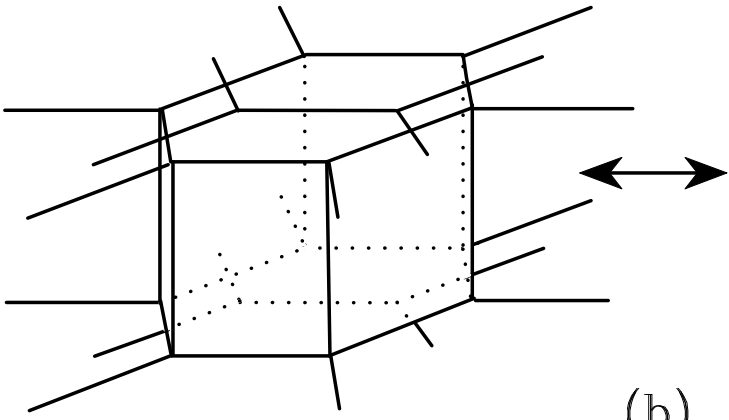
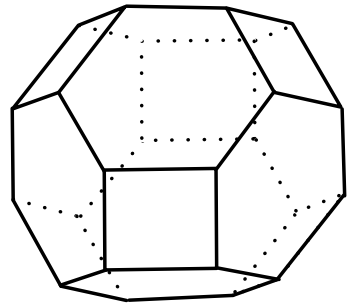
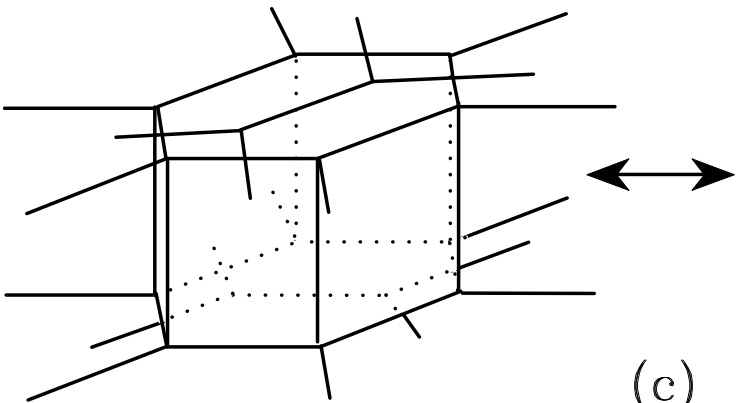
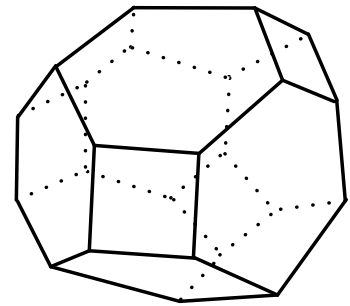
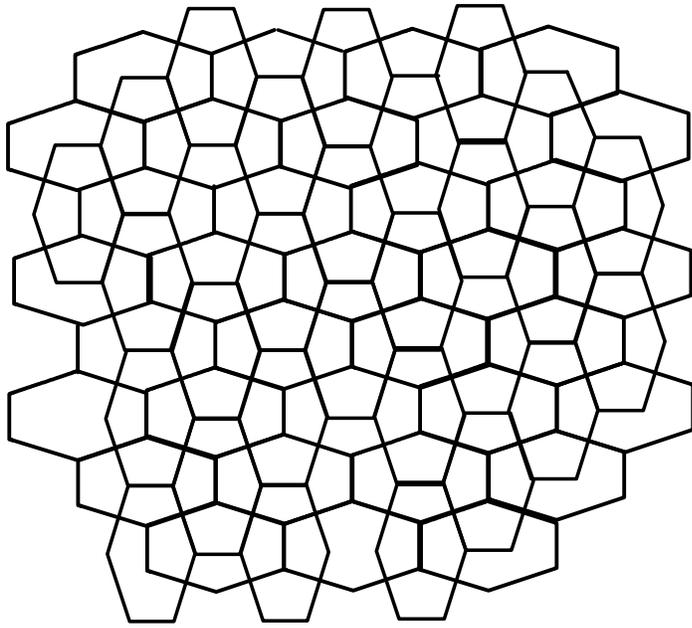
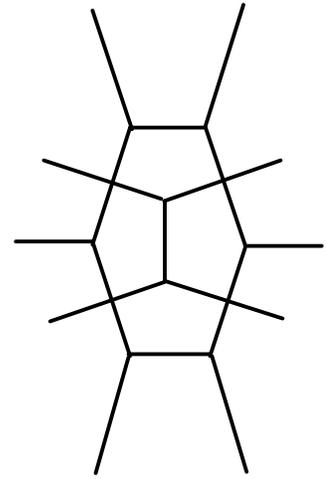
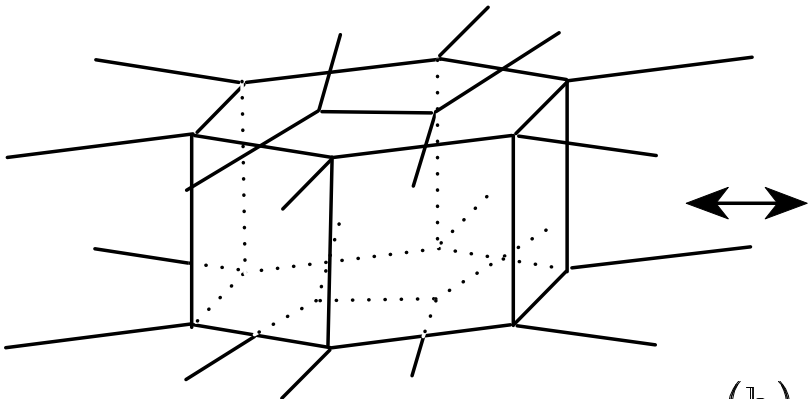
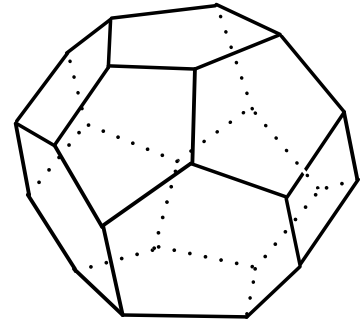
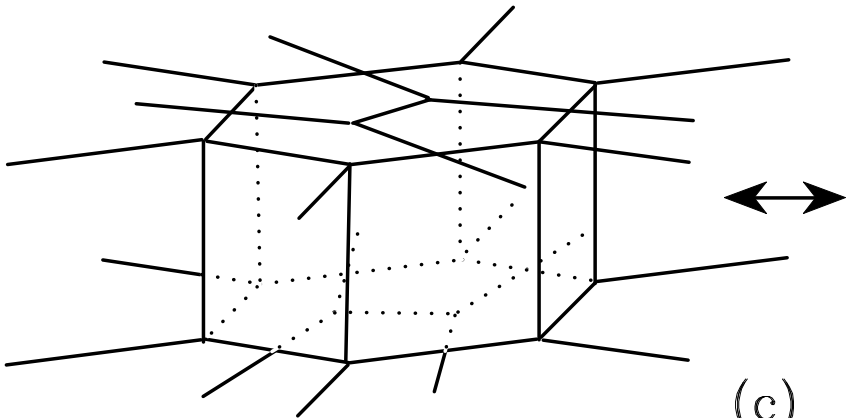
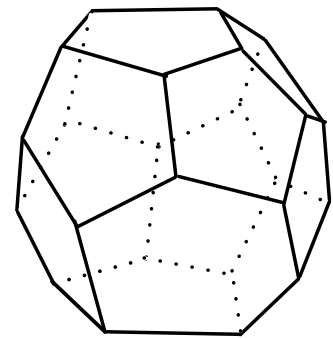
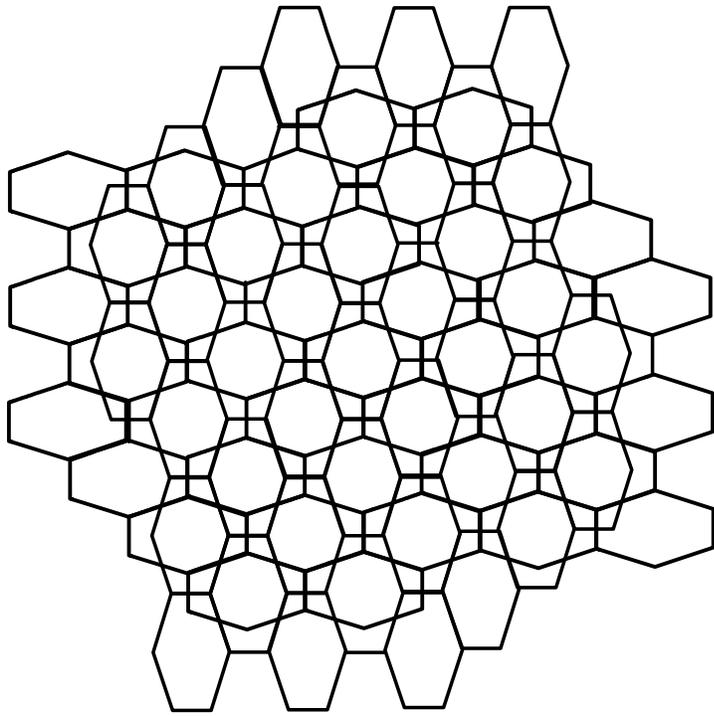
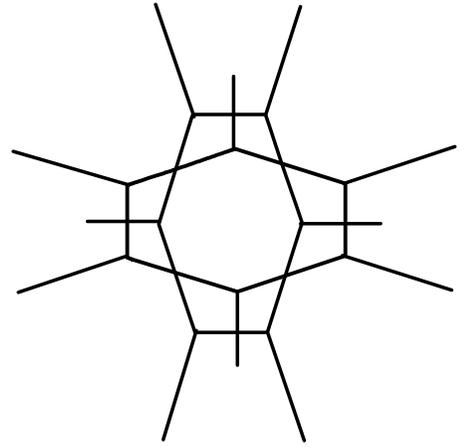
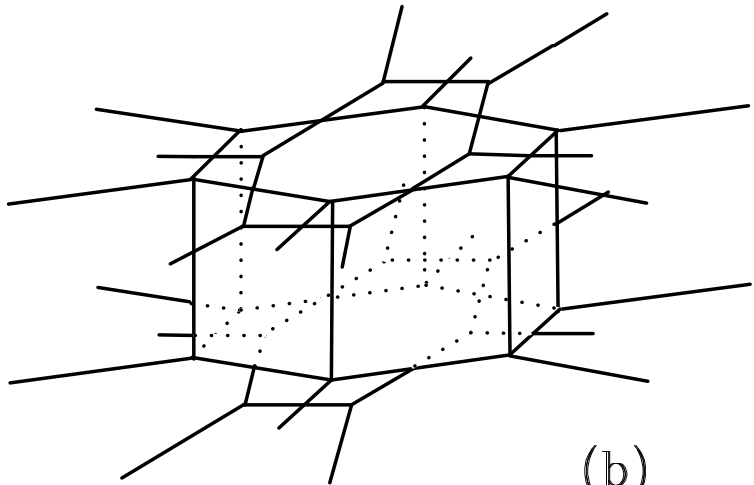
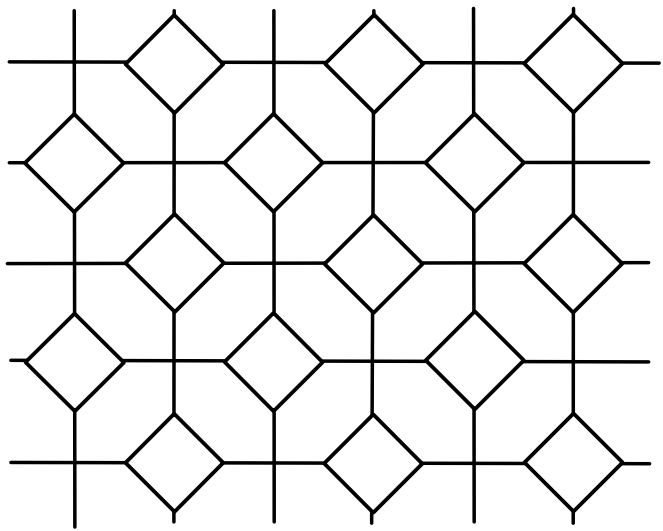
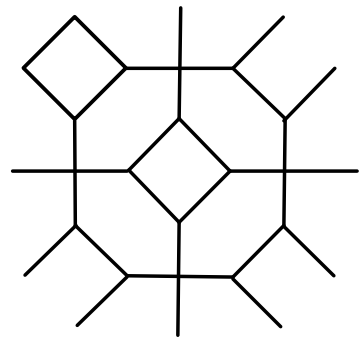
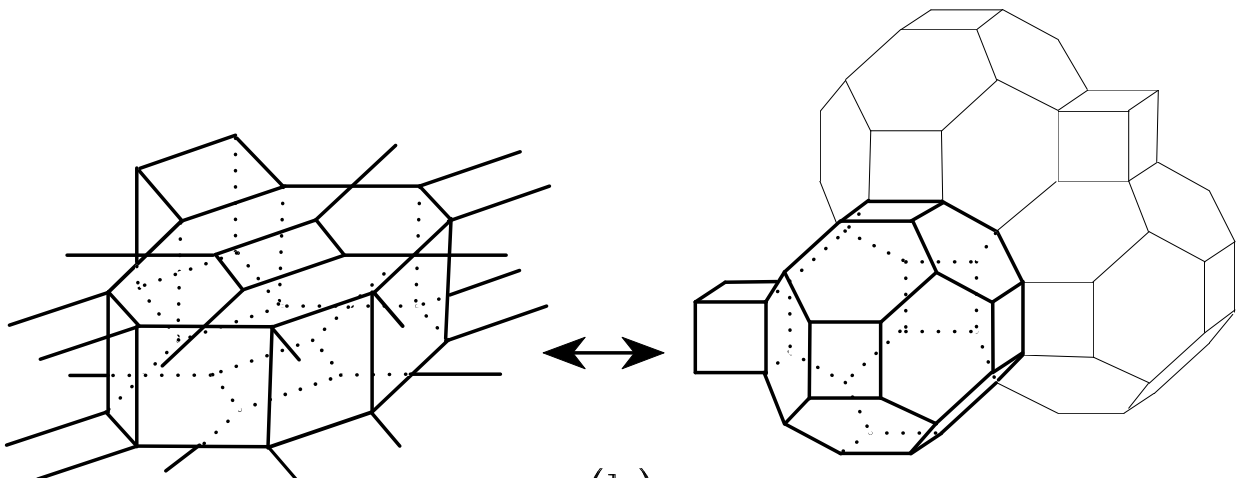
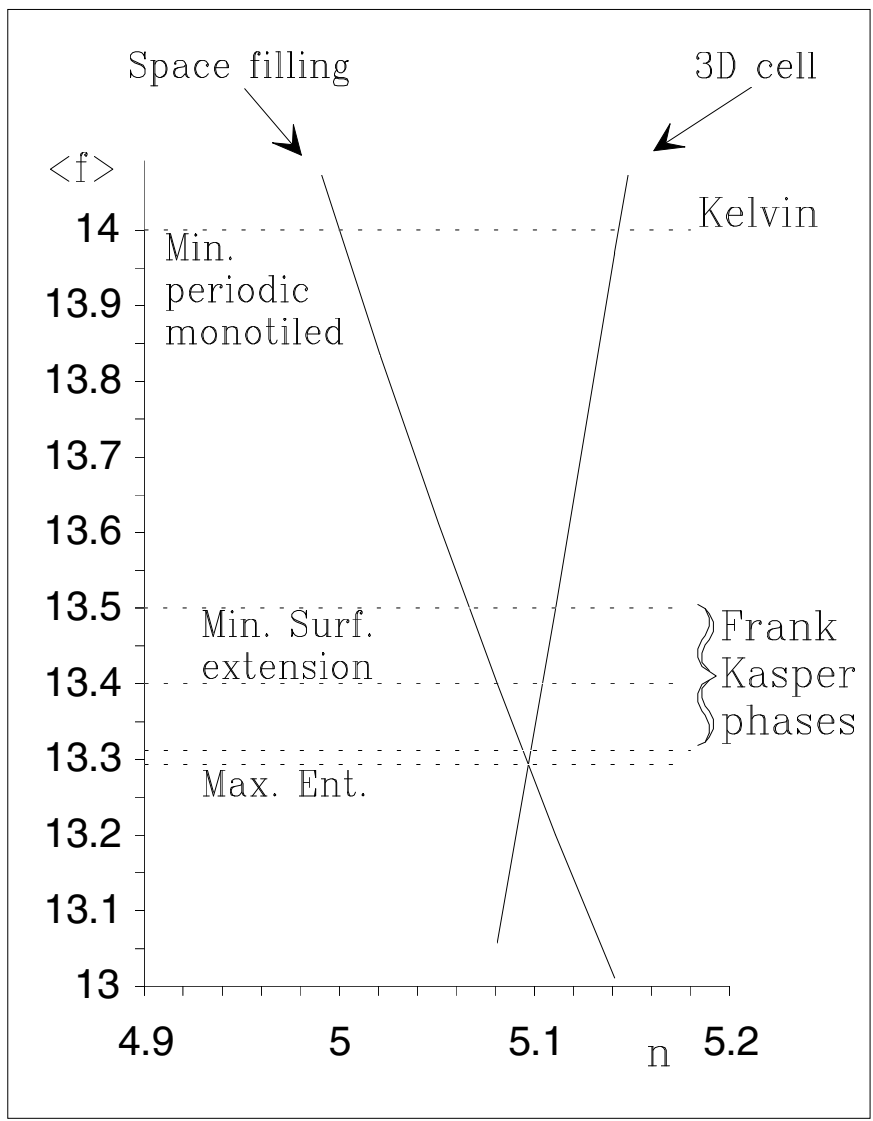
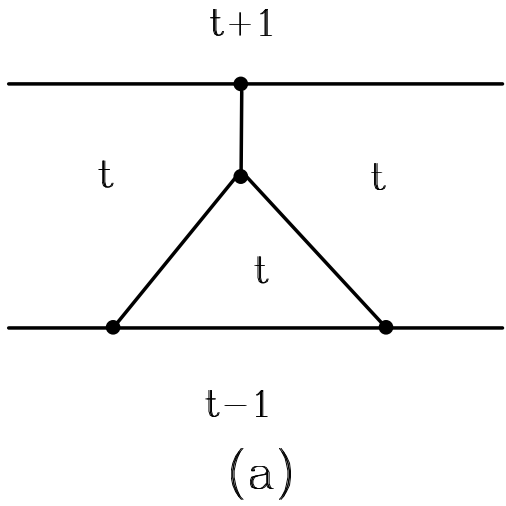
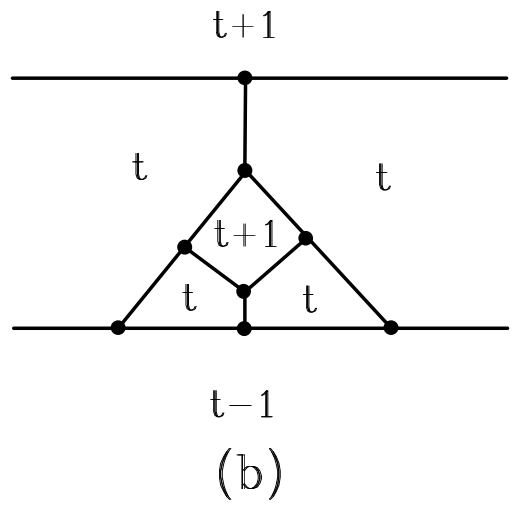
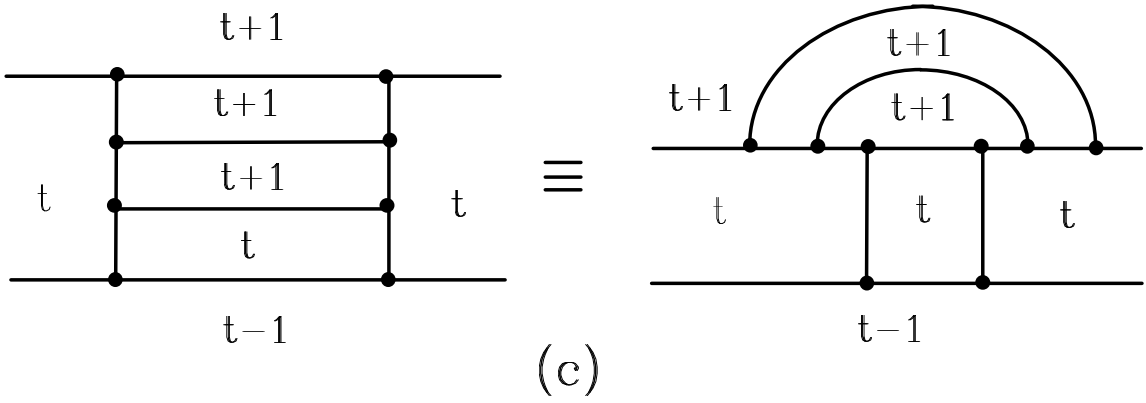
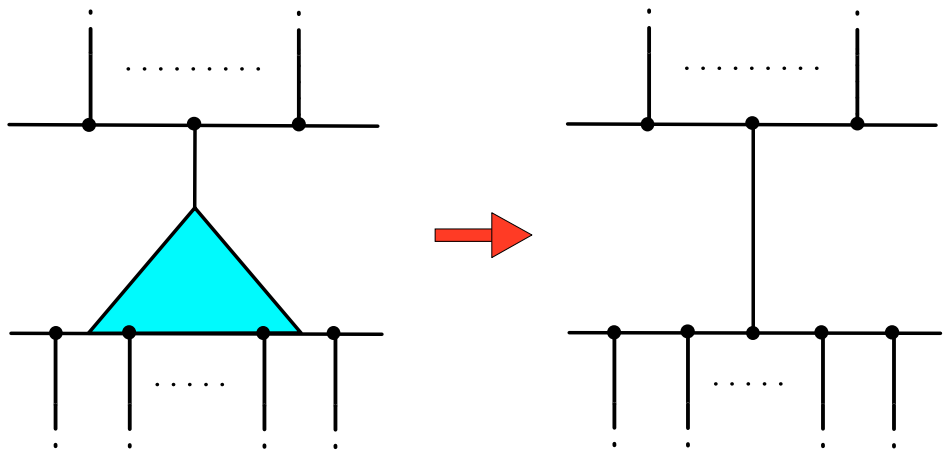
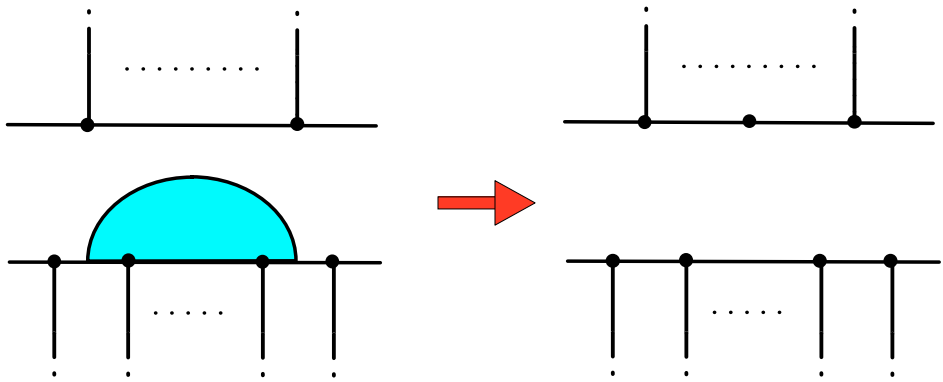
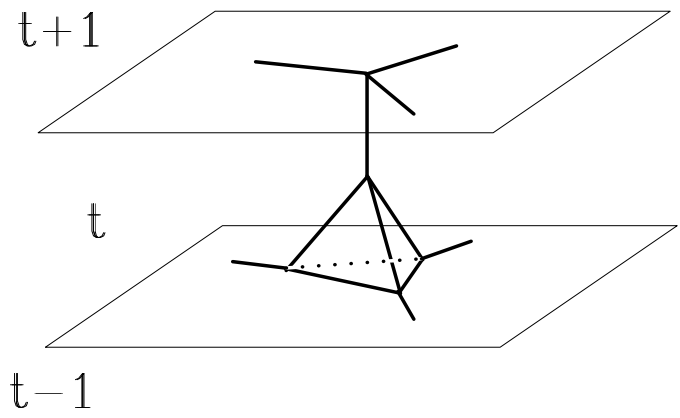
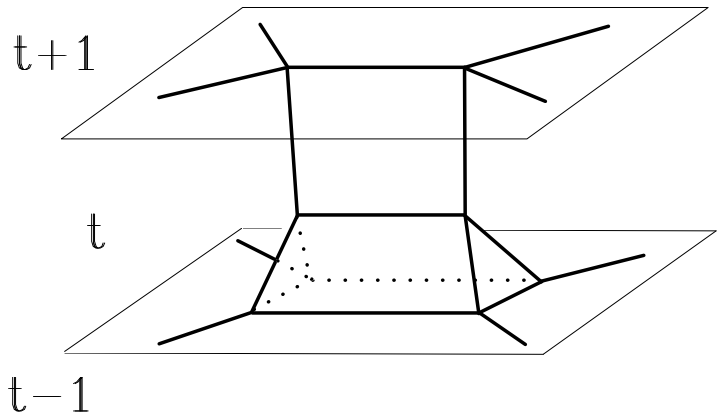
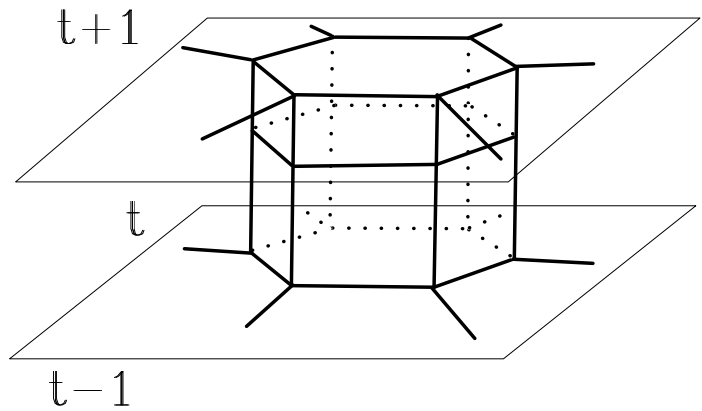
\end